\documentclass[pra,aps,twocolumn,nofootinbib,showpacs,floatfix,superscriptaddress,amsmath]{revtex4-1}
\usepackage{graphicx}
\newcommand {\fabs}[1] {\left| #1 \right|}
\newcommand {\degree}{^\circ}
\begin{document}
\title{Optimizing the catching of atoms or molecules in two-dimensional traps}
\author{V. P. Singh}
\email{vijay.singh@physnet.uni-hamburg.de}
\affiliation{Zentrum f\"ur Optische Quantentechnologien and Institut f\"ur Laserphysik, Universit\"at Hamburg, 22761 Hamburg, Germany}
\author{A. Ruschhaupt}
\email{aruschhaupt@ucc.ie}
\affiliation{Department of Physics, University College Cork, Cork, Ireland}
\begin{abstract}
Single-photon cooling is a recently introduced method to cool atoms and
molecules for which standard methods might not be applicable.
We numerically examine this method in a two-dimensional wedge trap as well as in
a two-dimensional harmonic trap. An element of the method is a small optical box with
``diodic'' walls which moves slowly through the external potential and catches
atoms irreversibly. We show that the cooling efficiency of the method can be improved by optimizing the trajectory of
this box.
\end{abstract}
\maketitle
\section{Introduction}
Recently, a new method has been introduced, called single-photon cooling
\cite{dudarev_2005,raizen_2009}, which allows to 
cool atoms and molecules which cannot be handled in a standard way.
The method is based on an atom diode or one-way barrier \cite{Andreas04,Raizen05}. An atom
diode is a device which can be passed by the atom only in one direction whereas the atom is
reflected if coming from the opposite direction. Such a device has been studied theoretically
\cite{Andreas06a,Andreas06b,Andreas06c,Andreas07}
and also experimentally implemented as a realization of a Maxwell demon 
\cite{thorn_2008, thorn_2009}. 

\begin{figure}[t]
\begin{center}
\includegraphics[angle=0,width=\linewidth]{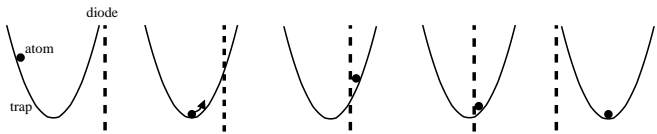}
\end{center}
\caption{\label{fig_intro_1}Schematic representation of the 
process for a single atom:
snapshots for different, increasing times (from left to right); the passing
direction of the diode is from left to right.}  
\end{figure}

The idea of cooling based on an atom diode is explained in
\cite{dudarev_2005} and \cite{Andreas06c}.
For illustration, we assume a simplified, idealized one-dimensional setting and
an ensemble of non-interacting atoms which are moving classically in a harmonic potential trap,
see Fig.~\ref{fig_intro_1}.
During the whole cooling process the atom diode is moved slowly through the trap
from right to left with a constant velocity. In addition, we assume that
the width of the diode is negligible.
An atom can cross the atom diode only
from left to right and after such a crossing the atom is trapped on the right-hand side of the diode.

For simplification let us look in the following at the process for a single atom
of the ensemble, see Fig.~\ref{fig_intro_1} and also \cite{Andreas06c}.
As the diode is moving much slower than the atom the diode captures every atom
near its classical turning point, i.e. the atom has very low kinetic energy
when it is caught.
If the atom crosses the diode from left to right then this is an irreversible process,
the atom cannot go back, i.e. the atom is caught between the diode and the trap wall.
Note that the implementations of an atom diode presented in
\cite{Andreas04,Andreas06a,Andreas06b,Andreas06c,Andreas07} are designed in
such a way that the internal state of the atom is changed during the passing of the diode but it will
be restored at the end in such a way that the atom is in the same internal state before and after
crossing the diode.
The diode, which is in principle a semi-penetrable barrier, behaves like a
wall for the captured atom and continues
moving to the left without changing its velocity.
During this the atom is bouncing off the slowly-moving diode and in such a way the kinetic energy of
the atom is reduced. Note that the total energy is not conserved during this process (a Hamiltonian
describing the system would be time-dependent).
So the diode transports the caught atom to the bottom of the potential without
increasing the kinetic energy of the atom, on the contrary, the kinetic energy
of the atom might be even more decreased.
This can be heuristically compared with a ball bouncing on a horizontal racket where the racket
is slowly moved down in the gravity field: Then the ball is not bouncing ``more'' if
the racket is arrived at the bottom.
Note that even if the atom has reached the bottom of the trap, diode still continues moving to the
left.
Finally, the atom is in the same external potential (this is
different for example from \cite{chu.1986}) and in 
the same internal state as it was initially, but its total energy has been reduced.

If the process acts on the whole ensemble of non-interacting
atoms -of a given temperature initially- then the energy of every atom
is reduced and therefore the ensemble is cooled. 

It is important to underline that -because of the
irreversible step- this cooling method is
fundamentally different from velocity reduction by collision with a moving
wall which is not ``real'' cooling \cite{ketterle.1992}.

Several variations of this cooling method have been proposed and applied in various experiments,
both for cooling atoms \cite{price_2007,price_2008,binder_2008,bannerman_2009,schoene_2010} and for
cooling molecules
\cite{narevicius_2009, shagam_2012}.

As a variant to the scheme shown in Fig.~\ref{fig_intro_1}, a small
optical dipole trap has been used in experiment \cite{price_2008}.
The small trap is constructed in the form of a square box \cite{price_2008}
and therefore we will simply call it ``box'' in this paper (not to be confused
with the external potential trap). The key idea is that this small box is
moved slowly though the external potential.
In an ideal setting, the walls of the box consist of atom diodes: if the atom
has crossed a wall of the box and the kinetic energy of the atom is lower
than the threshold energy of the ``diodic'' wall (called the threshold energy
of the box in the following) then the atom is irreversibly caught by the
box. Note that in the ideal case the caught atom is in the same internal state
as before the catching, see also
\cite{Andreas06a,Andreas06b,Andreas06c,Andreas07}.
In such a way, a similar effect as shown in
Fig.~\ref{fig_intro_1} may be achieved. 

\begin{figure}[t]
\begin{center}
(a)\includegraphics[angle=0,width=0.4\linewidth]{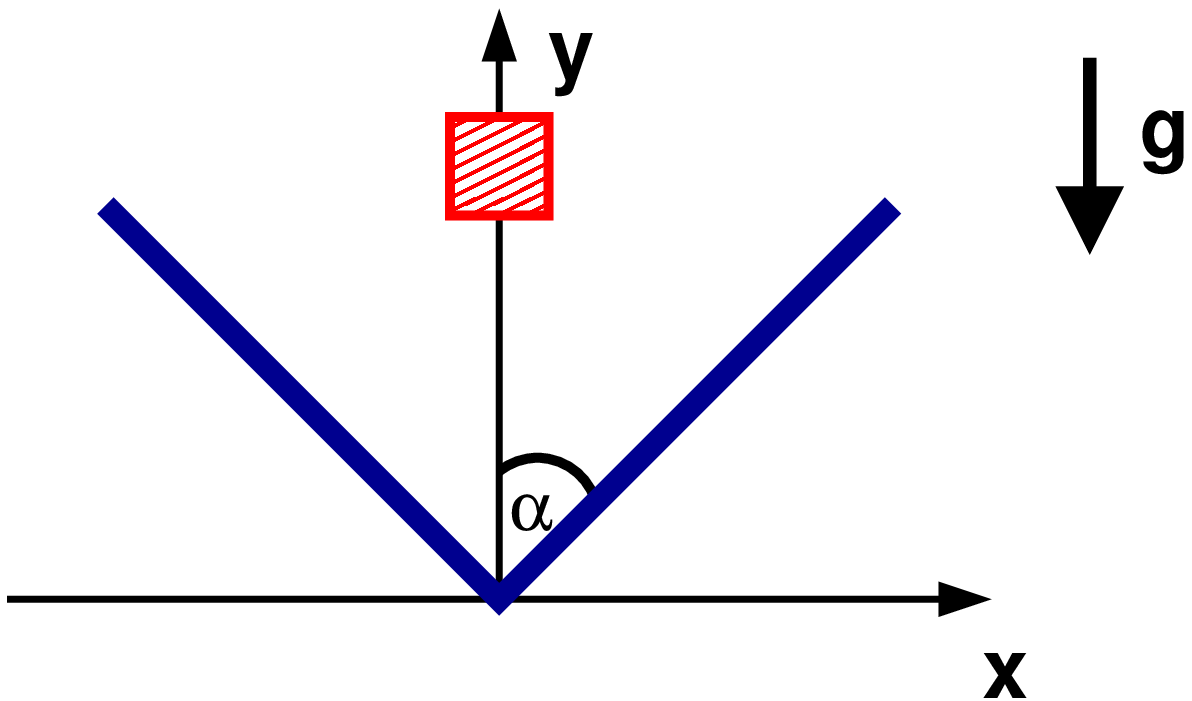}
(b)\includegraphics[angle=0,width=0.4\linewidth]{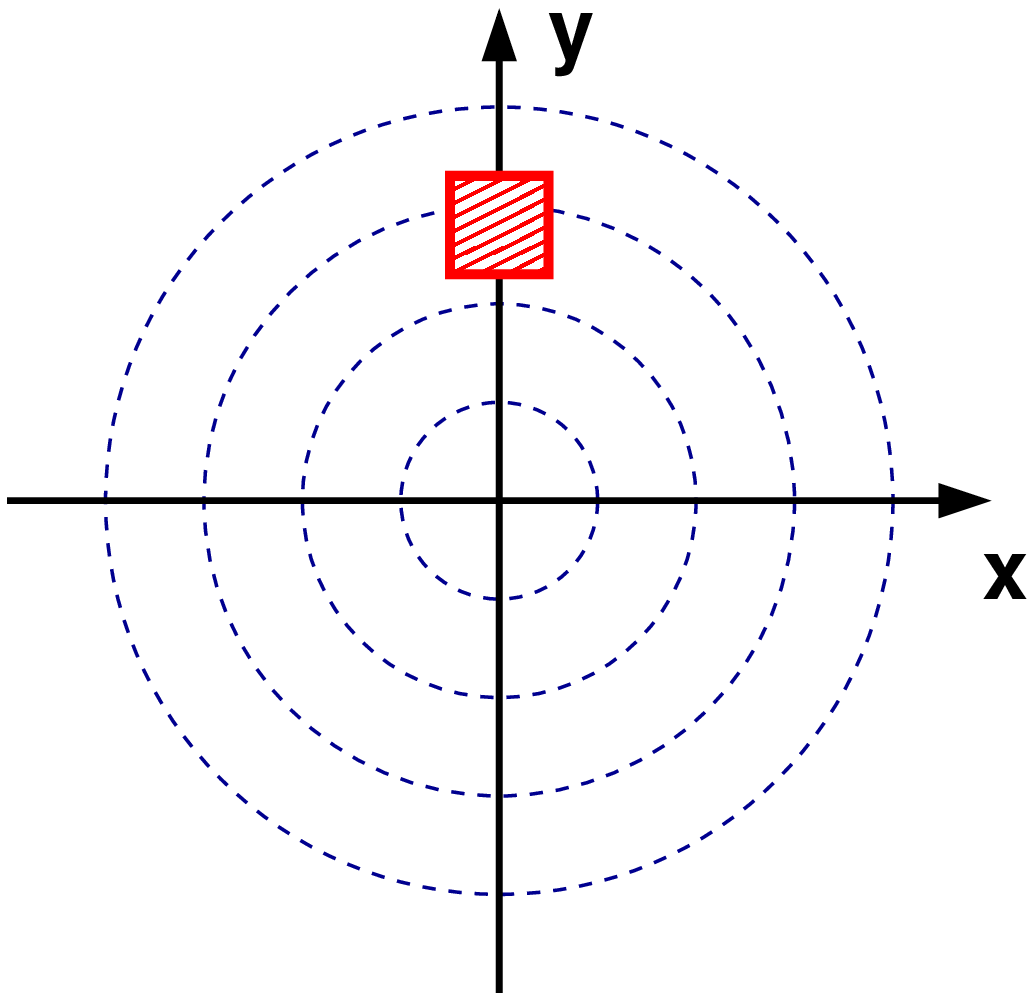}
\end{center}
\caption{\label{fig_intro_2}(Color online) Schematic representation of the two different
two dimensional traps considered: (a) wedge trap, (b) harmonic trap;
the red-filled squares indicate the ``box''.} 
\end{figure}

It is important that the box is moved in such a way that it traps
a maximal number of atoms. While this optimal trajectory of the box is
straightforward in a one-dimensional setting, this is not obvious
in higher dimensions. In \cite{choi_2010}, a
two-dimensional wedge trap has been examined. In that paper the box was
assumed to be at rest and the optimal position of the box was determined.
The main goal of our paper is to extend this work \cite{choi_2010} and
show that by a slowly moving box the cooling efficiency can be improved
compared to a box at rest. Therefore, we will examine
a simplified model of catching process in a two-dimensional wedge trap as
well as in a two-dimensional harmonic trap, see Fig.~\ref{fig_intro_2}. 
%
% ---------------- FIG. 1 BEGINS ----------------
% ---------------- END FIG. 1 ----------------
%
%
% ---------------- FIG. 2 BEGINS -------------------------
% --------------- Scheme: Two Trap Geometries ------------
% ---------------- END FIG. 2 ----------------------------
%

The paper is organized as follows: In Sect. \ref{sect2}, the model will be
explained in detail. We will then examine the setting of a two-dimensional wedge
trap in Sect. \ref{sect3} and of a two-dimensional harmonic trap in Sect. \ref{sect4}.
 
%-----------------------------------------------------------
%-----------------------------------------------------------
%-----------------------------------------------------------

\section{The Model\label{sect2}}

We examine two different two-dimensional trap geometries.
The first one is an idealized two-dimensional, symmetric wedge trap consisting
of perfectly reflecting walls, see Fig.~\ref{fig_intro_2}a. Gravity is
acting in $-y$ direction. The Hamiltonian
of a single atom in this trap is
\begin{equation}
H_W = \frac {1}{2m}(p_x^2 + p_y^2) + mg\,y + V(x,y),
\end{equation}  
where $V(x,y)=0$ if $y > 0, \fabs{x} < y \tan\alpha$ and $V(x,y)=\infty$ otherwise.
Moreover, $m$ is the mass of the atom and $g$ is the gravitational
acceleration at the earth's surface.
 
The second trap is a two-dimensional harmonic trap,
see Fig.~\ref{fig_intro_2}b. The Hamilton function of a single atom in this trap
is
\begin{equation}
H_H = \frac {1}{2m}(p_x^2 + p_y^2) + \frac {m}{2}\omega^2(x^2 + y^2),
\end{equation}
where $\omega$ is the frequency of the harmonic trap.

In both cases, we approximate the motion of the atoms
as a classical motion and in addition, we assume that the atoms inside the trap
do not interact with each other. So the trajectory of a single atom can be
calculated by solving Hamilton's equations
\begin{eqnarray}
\frac {\partial x}{\partial t} =  \frac {\partial H}{\partial p_x},\,
\frac {\partial y}{\partial t} =  \frac {\partial H}{\partial p_y},\,
\frac {\partial p_x}{\partial t} = - \frac {\partial H}{\partial x},\,
\frac {\partial p_y}{\partial t} = - \frac {\partial H}{\partial y}.\nonumber\\
\quad
\end{eqnarray}
The initial state of a single atom should be distributed concerning the
canonical probability distribution with temperature $T_i$ 
\begin{eqnarray}
\rho_i (x,y,p_x,p_y) = \frac{1}{Z} \exp\left[-\beta H (x,y,p_x,p_y)\right]\, ,
\label{rho0}
\end{eqnarray}
where $Z$ is the canonical partition function, $\beta = 1/(k_B T_i)$ and
$k_B$ is the Boltzmann constant. 

Our simplified model of the catching process is similar to and motivated by \cite{choi_2010}.
We model the ``box'', as a square box in
space with a width $2 w_B$ in $x$ and in $y$ direction, i.e. $w_B$ is the
half width of the box. 
The box can move freely throughout the trap.
We assume that the ``diodic'' walls of box are infinitely thin.
Moreover, the ``diodic'' walls of the box have a
certain threshold energy $E_B$ such that their ``diodic'' behavior breaks
down if the kinetic energy of the atom is above this threshold energy.
If the atom has crossed -at some time- the walls of the box and the kinetic
energy of the atom in the box rest frame is smaller than the threshold energy
$E_B$ then the atom is caught irreversibly by the
box for all times. If the kinetic energy of the atom is too large (larger
than $E_B$) then the
``diodic'' property of walls breaks down and the atom can escape the box
such that it might have a second chance later to be caught.

The algorithm is now the following:
For a single numerical run, we choose the initial state of a single
atom randomly with respect to the probability distribution Eq. (\ref{rho0}).
Then we calculate numerically the evolution of the atom until a final time
$t_f$ while always checking if the atom is in the region of the box during its
motion. If this is the case, we are checking if the kinetic energy of the atom
in the box rest frame is smaller than the box threshold energy $E_B$. If this
is true, then the atom is caught and the run is finished, otherwise the motion
of the atom continues and it may get trapped later.
We repeat such a single run $N_i=10^6$ times, get the total number of trapped atoms
$N_B$ and finally the relative number of trapped atoms, i.e. the fraction $F=N_B/N_i$.

The goal of this paper is to optimize this fraction $F$ by varying the
trajectory of the box in the trap. As we will see now, from this follows 
also an optimization of the cooling efficiency if the initial conditions and the
box parameters are fixed.

One can define a cooling efficiency as the compression of the phase-space density. 
Phase-space density in the context of cooling can be defined as the number of atoms in a box with
sides of one ``thermal de Broglie wavelength'' \cite{Townsend95}. The ``thermal de
Broglie wavelength'' $\lambda$ can be defined as
\begin{eqnarray}
\lambda = \frac{\hbar\sqrt{2\pi}}{\sqrt{m k_B T}},
\end{eqnarray}
where $m$ is the mass of the
atom. The phase-space density in a two-dimensional setting
is then defined as $\mu = n\lambda^2$, where $n$ is the spatial density
of atoms. The relative change in phase-space density is therefore
\begin{equation}
\frac {\mu_f}{\mu_i} = \frac {n_f}{n_i} \frac {T_i}{T_f},
\end{equation}
where $\mu_i$ is the initial and $\mu_f$ is the final phase-space density.
$T_i$ and $T_f$ are the initial and final  temperature, respectively.

The initial distribution will be a canonical one with temperature $T_i$.
The initial spatial density $n_i$ is strictly speaking not constant in
space. Nevertheless, we set approximately $n_i = N_i/A_i$, where
$N_i$ is the initial number of atoms. To get $A_i$, we fix a small $\varepsilon>0$
and define the initial region $A_i$ such that the atom is initially in $A_i$ with
probability $1-\varepsilon$.
The final spatial density is set as $n_f = N_B/A_B$, where $N_B$ is the number of trapped atoms
and $A_B = 4 w_B^2$ is the area of the box.

The cooling efficiency $\eta$ is then defined as
\begin{equation} 
\eta = \log_{10}\left( \frac {\mu_f}{\mu_i} \right)
= \log_{10}\left( \frac {N_B}{N_i} \frac {A_i}{A_B}
\frac {T_i}{T_f}\right).
\end{equation} 
Here $E_i = k_B T_i$. For simplicity, the final temperature is approximated
by the threshold energy of the box, i.e. $T_f = E_{B}/k_B$.
The cooling efficiency depends on the fraction of
trapped atoms $F=N_B/N_i$, the ratio  of the area of the thermal
system to the area of the box,
$A_i/A_B$, and the ratio of the initial to final energy, $E_i/E_B$:
\begin{equation} 
\eta = \log_{10}\left( F \frac {A_i}{A_B}
\frac {E_i}{E_B}\right).
\end{equation} 
If the initial conditions and the
box parameters are fixed, it will be therefore sufficient to optimize
the fraction of trapped atoms $F$. The goal of the rest of the paper will be
to optimize this fraction by choosing different box trajectories.

%-----------------------------------------------------------
%-----------------------------------------------------------
%-----------------------------------------------------------

\section{Box trajectories in a wedge trap\label{sect3}}
 
First we study the wedge trap, see Fig.~\ref{fig_intro_2}a.
According to Eq. (\ref{rho0}), the initial canonical distribution is
given by 
\begin{eqnarray}
\rho_{i,W} (x,y,p_x,p_y) = 
\frac {mg^2 \beta ^3 \chi_{[-y \tan \alpha,\, y \tan \alpha]}(x)}{4\pi \tan \alpha}\nonumber\\
\times \exp\left\{- \beta \left[({p_x}^2 + {p_y}^2)/(2m) +
  mgy\right]\right\}\, ,
\end{eqnarray} 
where $\chi_J (x)$ is the indicator function with $\chi_J (x)=1$ if $x \in J$
and $\chi_J (x)=0$ if $x \notin J$.
It is convenient to define a characteristic length $l$, a
characteristic velocity $\nu$ and a characteristic time $\tau$ by
\begin{eqnarray}
l = \frac{k_B T_i}{mg}, \nu = \sqrt{\frac{k_B T_i}{m}}, \tau =
\sqrt{\frac{k_B T_i}{mg^2}}\, .
\end{eqnarray}
In the rest of the paper we are assuming$~^{87}{\rm Rb}$ atoms,
i.e. $m=\mbox{mass}(^{87}{\rm Rb})=1.44316\times 10^{-25}\, {\rm kg}$.
The initial temperature should be $T_i = 100 \,\mu{\rm K}$ and $g=9.78 \, {\rm m/s^2}$ (equator). The characteristic
values are then
\begin{eqnarray}
l = 978\, \mu{\rm m},\quad 
\nu = 9.78\, {\rm cm/s},\quad
\tau = 10\, {\rm ms}.
\end{eqnarray}

\subsection{Box at rest}

\begin{figure}{}
\begin{center}
(a) \includegraphics[width=0.92\linewidth]{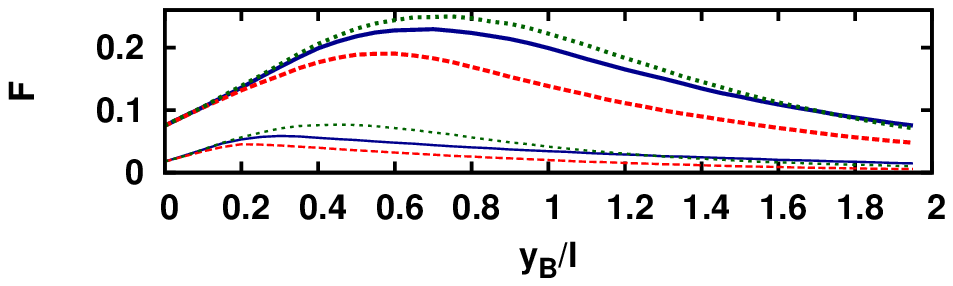}

(b) \includegraphics[width=0.85\linewidth]{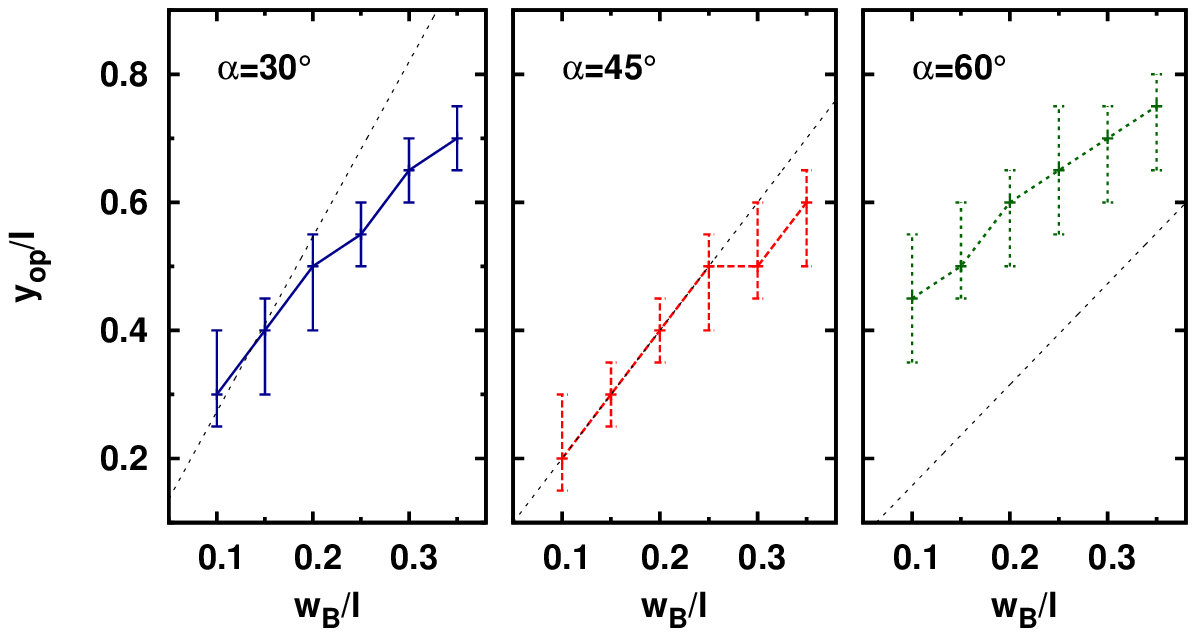}
\end{center}
\caption{(Color online) Wedge trap with the box at rest at $x_B = 0$.
(a) Fraction $F$ of trapped atoms versus box positions $y_B$:
$\alpha=30^\circ,w_B/l=0.1$ (solid blue line);
$\alpha=45^\circ,w_B/l=0.1$ (dashed red line);
$\alpha=60^\circ,w_B/l=0.1$ (dotted green line);
$\alpha=30^\circ,w_B/l=0.35$ (thick solid blue line);
$\alpha=45^\circ,w_B/l=0.35$ (thick dashed red line);
$\alpha=60^\circ,w_B/l=0.35$ (thick dotted green line).
(b) Optimal position $y_{op}$ of the box at rest for different
wedge angles (blue, red resp. green plus signs connected with lines),
the black dots correspond to the straight line
$y_{op}=w_B (1+\tan\alpha)/\tan\alpha$, see text for more details.
$t_f/\tau = 20$, $E_B/E_i=0.1$.
\label{fig_wedge_1}}
\end{figure}

The center of the box is moving with a trajectory $(x_B(t),y_B(t))$.
As a reference case we first assume that the box center is placed at $x_B=0$ at rest,
i.e. its velocity is $v_{B,x}=v_{B,y}=0$.
Fig.~\ref{fig_wedge_1}a shows the
trapping fraction $F$ versus different box positions $y_B$ for different
combinations of wedge angle $\alpha$ and box half width $w_B$.

The optimal height $y_{op}$ which gives the maximal fraction of trapped atoms for
different box half widths $w_B$ is shown in Fig.~\ref{fig_wedge_1}b. The error bars
are defined by the range in which the maximal trapping fraction $F$ decays by
an amount of $1/\sqrt{N_i}$, where $N_i$ is the number of particles used in
the numerical simulation as defined above.

% ---------------- Fig. 3 --------------------------------------
% ------------ Wedge: Box at rest ------------------------------

% ---------------- Fig. 3 --------------------------------------

\subsection{Box with linear motion}

We want to examine if a moving box can produce a higher fraction $F$ of
trapped atoms than the box at rest. 
First, we consider a linear motion of the box given by
\begin{eqnarray}
\begin{array}{rcl}
x_B(t) &=& v_{B,x} (t-t_f/2)\, , \\
y_B(t) &=& v_{B,y} (t-t_f/2) + y_{op}\, .
\end{array}
\label{wtrajlinear} 
\end{eqnarray}
Here $y_{op}$ is chosen to be equal to the optimal value for the
box at rest shown in Fig.~\ref{fig_wedge_1}b (for the corresponding wedge
angle and box half width).
Because of mirror symmetry ($x \leftrightarrow -x$) of the setting we can
restrict to the case $v_{B,x}>0$.

\begin{figure}{}
(a) \includegraphics[width=0.8\linewidth]{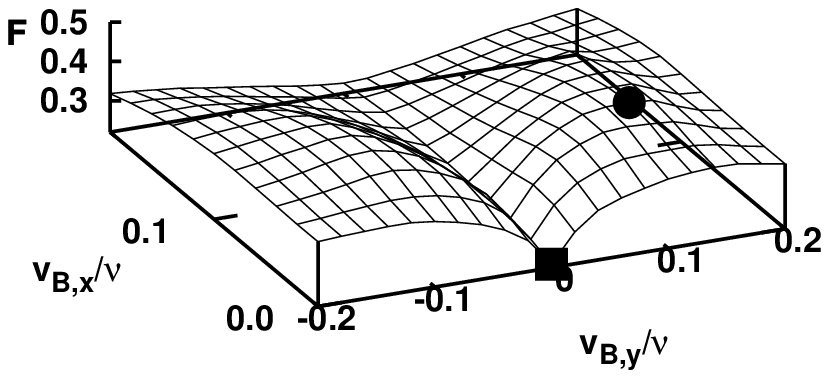}

(b) \includegraphics[width=0.8\linewidth]{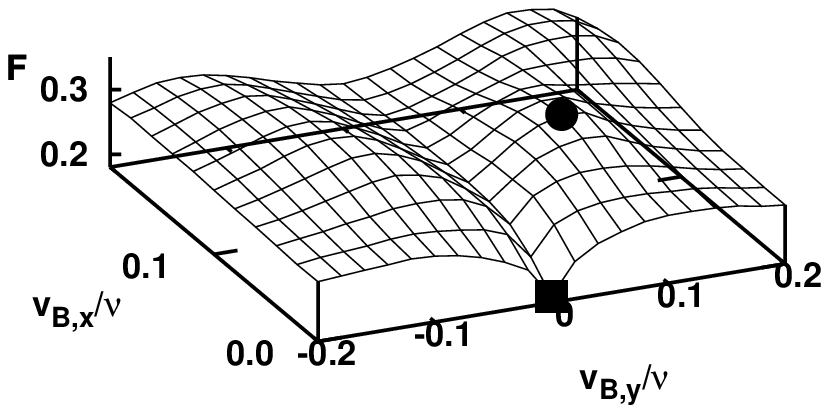}

(c) \includegraphics[width=0.8\linewidth]{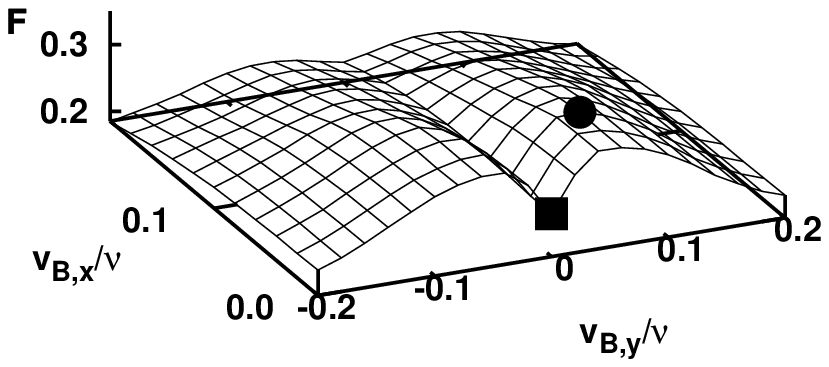}
\caption{Wedge trap. Box moving with the linear trajectory Eq.~(\ref{wtrajlinear}).
Fraction $F$ of trapped atoms versus box velocities:
(a) wedge angle $\alpha=30^\circ$, $y_{op}/l=0.7$;
(b) wedge angle $\alpha=45^\circ$, $y_{op}/l=0.6$;
(c) wedge angle $\alpha=60^\circ$, $y_{op}/l=0.75$.
$t_f/\tau = 20$, $E_B/E_i=0.1$, $w_B/l=0.35$.
The square symbols mark the maximal fraction for a
box at rest, the dots mark the maximal fraction for a linear moving box.
\label{fig_wedge_2}}
\end{figure}

Fig.~\ref{fig_wedge_2} shows the resulting fraction $F$ versus different box
velocities for different wedge angles. The box half width is fixed at
$w_B/l=0.35$.
The square symbols mark the maximal fraction $F$ which can be achieved with a
box at rest while the dots mark the maximal fraction which can be achieved with a
box moving with the trajectory Eq.~(\ref{wtrajlinear}). It is clearly seen that
the linear moving box can capture a larger fraction $F$ of atoms than the box
at rest (for all the three different wedge angles $\alpha=30\degree,45\degree$ and $60\degree$).

The optimal velocity in the first case $\alpha=30\degree$
(Fig.~\ref{fig_wedge_2}a) is
\begin{eqnarray}
\begin{array}{rcccl}
v_{B,x}/\nu &=& 0.06 &\approx& 0.13 \,\sin 30\degree\, ,\\
v_{B,y}/\nu  &=& 0.12 &\approx& 0.13\,\cos 30\degree\, .
\end{array}\label{vopt30}
\end{eqnarray}
In the case $\alpha = 45\degree$ (Fig.~\ref{fig_wedge_2}b), we get for the optimal velocity
\begin{eqnarray}
\begin{array}{rcccl}
v_{B,x}/\nu &=& 0.08 &\approx& 0.11 \,\sin 45\degree\, ,\\
v_{B,y}/\nu &=& 0.08 &\approx& 0.11\,\cos 45\degree\, .
\end{array}\label{vopt45}
\end{eqnarray}
It is important to notice from Eqs. (\ref{vopt30}) and (\ref{vopt45})
that we get in both cases the maximal fraction if the box is approximately
moving parallel to one wedge side.

This is different in the case $\alpha=60\degree$ (Fig.~\ref{fig_wedge_2}c). The optimal velocities
are now
\begin{eqnarray}
\begin{array}{rcccl}
v_{B,x}/\nu &=& 0.04 &\approx& 0.072 \,\sin 34\degree\, ,\\
v_{B,y}/\nu &=& 0.06 &\approx& 0.072\,\cos 34\degree\, ,
\end{array}\label{vopt60}
\end{eqnarray}
i.e. the box motion is not parallel to the wedge side in this case.
Nevertheless, the linear moving box traps in all cases more atoms than
the corresponding box at rest.

Therefore, the box moving with the linear trajectory enhances the
trapping fraction than the box at rest.
In the following, we will study the linear moving box in more detail to see
if this result remains true for other box half widths.

% ---------------- Fig. 4 --------------------------------------
% --------- Wedge: Linear Motion (3D) --------------------------
% ---------------- Fig. 4 --------------------------------------

% ---------------- Fig. 5 --------------------------------------
% --------- Wedge: Linear Motion -------------------------------
\begin{figure}{}
\begin{center}
\includegraphics[width=0.4\linewidth]{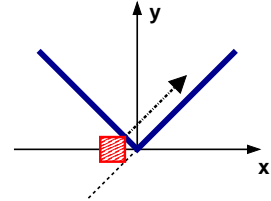}
\end{center}
\caption{(Color online) Schematic representation of the box
  trajectories Eqs. (\ref{wtrajnum}) and (\ref{wtrajanal}), the initial
  position of the box is shown, see text for further details.
\label{fig_wedge_0}}
\end{figure}
% ---------------- Fig. 5 --------------------------------------

\begin{figure}{}
\begin{center}
(a) \includegraphics[width=0.9\linewidth]{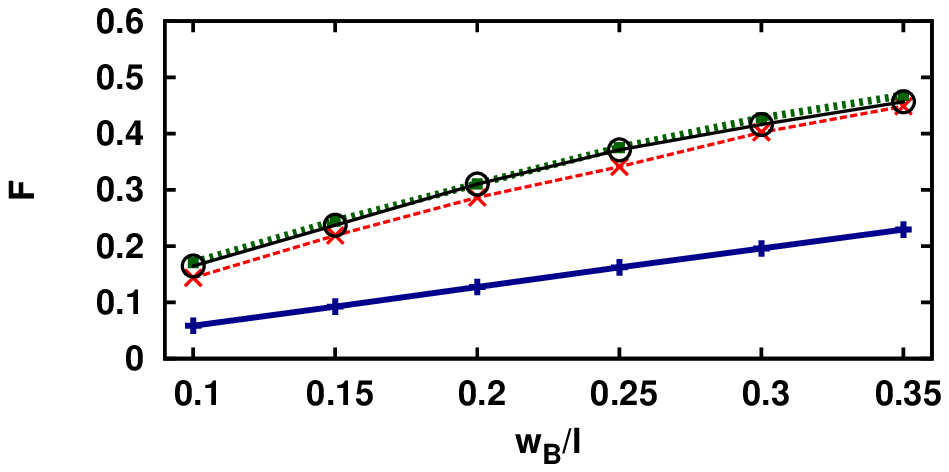}

(b) \includegraphics[width=0.9\linewidth]{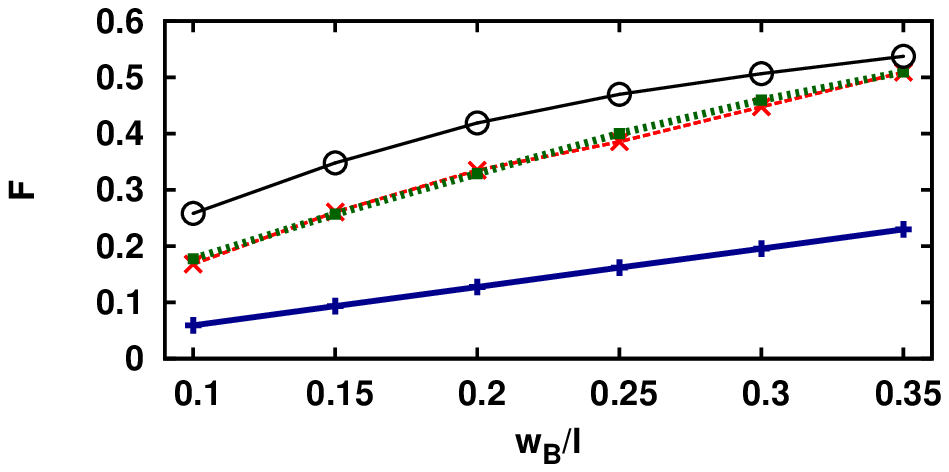}
\end{center}
\caption{(Color online) Wedge trap, $\alpha=30^\circ$. Fraction $F$ of trapped atoms versus
  box half widths $w_B$: (a) $t_f/\tau = 20$, (b) $t_f/\tau =
  40$; different
  box trajectories:
box at rest (plus signs connected by thick blue line),
box moving with trajectory Eq. (\ref{wtrajnum}) (crosses connected by red dashed line),
box moving with analytical trajectory Eq. (\ref{wtrajanal}) (dots connected by
thick green dotted line),
box moving with wriggle trajectory Eq. (\ref{wtrajwriggle})
(circles connected by black solid line).
$v/\nu = 0.13$, $y_{W0}/l=2.0$, $\omega_W\tau=0.25$, $E_B/E_i=0.1$. 
\label{fig_wedge_3}}
\end{figure}

\begin{figure}{}
\begin{center}
(a) \includegraphics[width=0.9\linewidth]{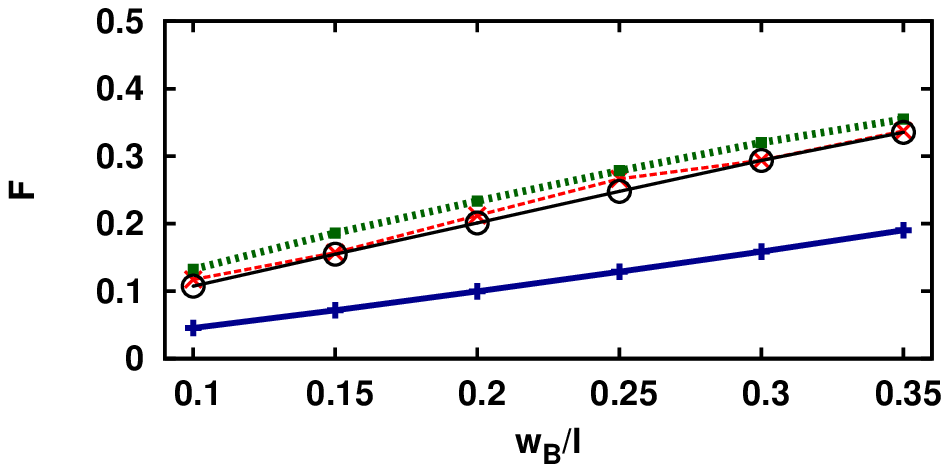}

(b) \includegraphics[width=0.9\linewidth]{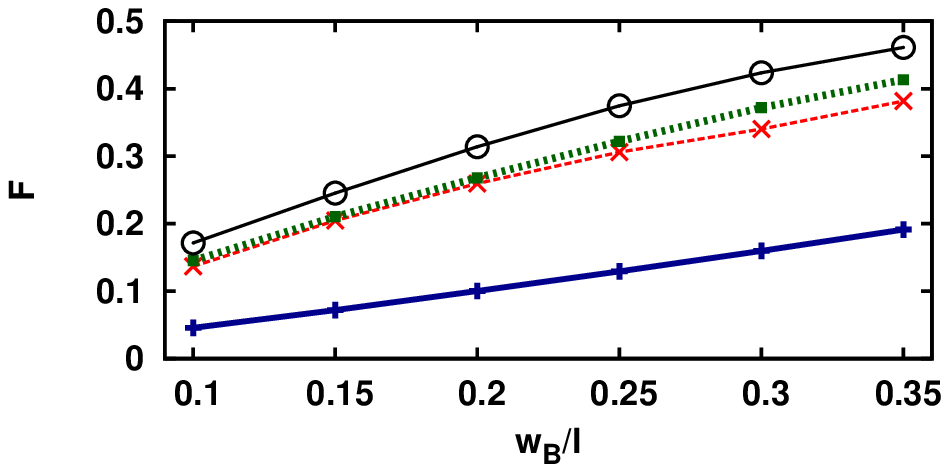}
\end{center}
\caption{(Color online) Wedge trap, $\alpha=45^\circ$. Fraction $F$ of trapped atoms versus
  box half widths $w_B$: (a) $t_f/\tau = 20$, (b) $t_f/\tau = 40$.
$v/\nu = 0.11$, $y_{W0}/l=1.5$, $\omega_W\tau=0.2$; see Fig.~\ref{fig_wedge_3} for more details.
\label{fig_wedge_4}}
\end{figure}

We first look at the cases of a wedge angle $\alpha=30\degree$ and
$\alpha=45\degree$, respectively.
Figs.~\ref{fig_wedge_3} and \ref{fig_wedge_4} show the fraction $F$ of trapped
atoms versus different box half widths $w_B$. Figs.~\ref{fig_wedge_3}a
and \ref{fig_wedge_4}a correspond to the final time $t_f/\tau =
20$ while Figs.~\ref{fig_wedge_3}b and \ref{fig_wedge_4}b
correspond to $t_f/\tau = 40$.

The results for a box at rest ($x_B=0, v_{B,x}=v_{B,y}=0$) are shown as a reference
case for $t_f/\tau = 20$ as well as for $t_f/\tau = 40$ (plus signs connected
by thick blue line) with the box coordinate $y_B=y_{op}$ and $y_{op}$
shown in Fig.~\ref{fig_wedge_1}b (which is optimal for $t_f/\tau = 20$).
It can be seen in Figs.~\ref{fig_wedge_3} and \ref{fig_wedge_4} 
that for a box at rest the fraction does not change significantly if the final
time $t_f$ increases.

Motivated by the previous results, we are again consider a box which moves
linearly in direction of the right wedge side, i.e. its velocity is $v_{B,x}=v \sin\alpha, v_{B,y}=v \cos\alpha$.
The box trajectory should cross the $y$-axis at $y_{op}$, where $y_{op}$ is
again the position shown in Fig.~\ref{fig_wedge_1}b. 
Nevertheless, the box should now start at time $t=0$ directly outside the wedge trap,
see Fig.~\ref{fig_wedge_0}. Note that therefore the $y$ axis is no longer
crossed at time $t_f/2$ by the box center.
The resulting trajectory is
\begin{eqnarray}
\begin{array}{rcl}
x_B (t) &=& v \sin\alpha\,t - \frac{1}{2} \left(w_B + (w_B +  y_{op})\tan\alpha\right)\, ,\\
y_B (t) &=&  v \cos\alpha\,t - \frac{1}{2} \left(w_B - y_{op} + w_B \cot\alpha\right)\, .
\end{array}
\label{wtrajnum}
\end{eqnarray}
The results for such a moving box for $t_f/\tau = 20$ as well as for $t_f/\tau = 40$ are shown
in Figs.~\ref{fig_wedge_3}-\ref{fig_wedge_4} (crosses connected by red dashed line).
It can be seen that the fraction $F$ is -in all cases- much larger for such a
linearly moving box compared to the box at rest.

However, these box trajectories still depend on the values $y_{op}$ which
have to be obtained by numerical optimization. The goal is now to find an
analytical approximation for $y_{op}$.
Therefore, we are considering a box moving parallel to the right wedge
side while touching the wedge side with its right, lower corner
(see also Fig.~\ref{fig_wedge_0}). Then we get the analytical value
$y_{op} = w_B (1 + \tan\alpha)/\tan\alpha$.
Note that these values for $y_{op}$ are also plotted in
Fig.~\ref{fig_wedge_1}b and we can see that this is also a rough approximation
for the numerical determined optimal position $y_{op}$.
This ``analytical'' box trajectory is now
\begin{eqnarray}
\begin{array}{rcl}
x_B (t) &=& v \sin\alpha\,t - w_B (1 + \tan\alpha)\, ,\\
y_B (t) &=&  v \cos\alpha\,t \, .
\end{array}
\label{wtrajanal}
\end{eqnarray}
The resulting fraction $F$ using the box trajectory
Eq. \ref{wtrajanal} can be seen in Figs.~\ref{fig_wedge_3}-\ref{fig_wedge_4} 
(dots connected by thick green dotted line). We find that the box traps even more
atoms than with the trajectory Eq. (\ref{wtrajnum}) considered in the previous
paragraph and it has the advantage that no numerical determined value of
$y_{op}$ is required.

\begin{figure}{}
\begin{center}
(a) \includegraphics[width=0.9\linewidth]{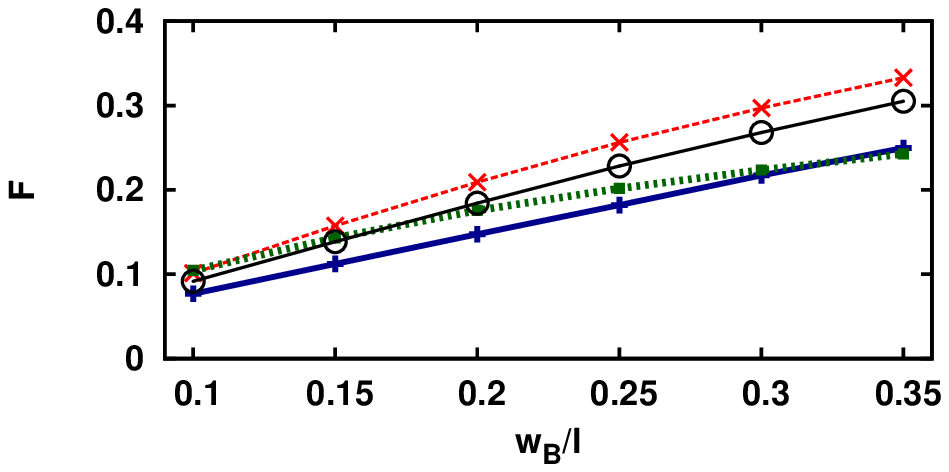}

(b) \includegraphics[width=0.9\linewidth]{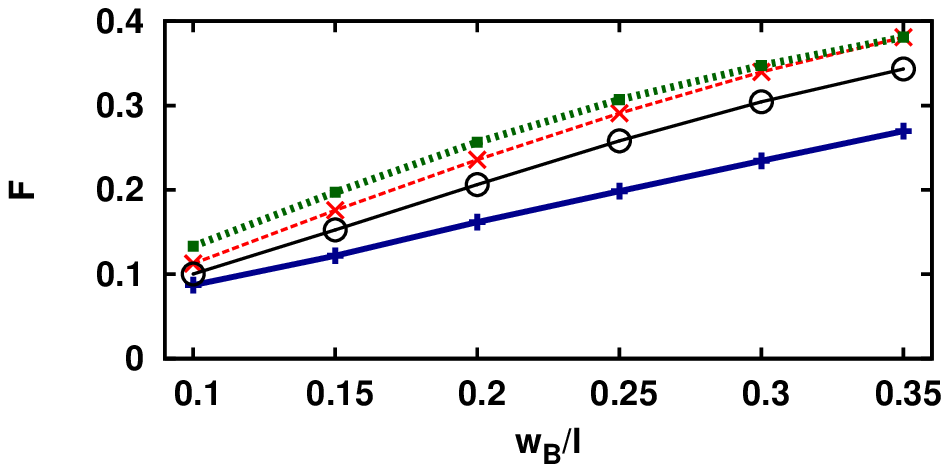}
\end{center}
\caption{(Color online) Wedge trap, $\alpha=60^\circ$. Fraction $F$ of trapped atoms versus
  box half widths $w_B$: (a) $t_f/\tau = 20$, (b) $t_f/\tau = 40$.
$v/\nu = 0.11$, $y_{W0}/l=1.5$, $\omega_W\tau=0.2$;
see Fig.~\ref{fig_wedge_3} for more details.
\label{fig_wedge_5}}
\end{figure}

For completeness, the fraction $F$ versus the box half widths for $\alpha=60\degree$ is shown in
Fig.~\ref{fig_wedge_5}. The results are shown for a box at rest (plus signs
connected by a thick blue line) and a box moving linearly with
trajectory Eq. (\ref{wtrajnum}) as well as Eq. (\ref{wtrajanal}). The value of $y_{op}$
in Eq. (\ref{wtrajnum}) is chosen from Fig.~\ref{fig_wedge_1}b.
We see that the moving box catches more atoms than
the box at rest. In contrast to the cases $\alpha=30\degree$ and
$\alpha=45\degree$, the analytical box trajectory Eq. (\ref{wtrajanal}) is here a good choice
only for $t_f/\tau=40$.

% ---------------- Fig. 6 --------------------------------------
% --------- Wedge: Linear Motion -------------------------------
% ---------------- Fig. 6 --------------------------------------

% ---------------- Fig. 7 --------------------------------------
% --------- Wedge: Linear Motion -------------------------------
% ---------------- Fig. 7 --------------------------------------

% ---------------- Fig. 8 --------------------------------------
% --------- Wedge: Linear Motion -------------------------------
% ---------------- Fig. 8 --------------------------------------

\subsection{Box moving with wriggled trajectory}

We want to underline that a linear trajectory might not be
optimal. However, the advantage of a linear motion of the box is that it
might be easier to implement experimentally a linear motion than a more
complicated motion of the box.

Nevertheless, as an example, we shall also try a different trajectory of the
box, a wriggled one, given by
\begin{eqnarray}
\begin{array}{rcl}
x_B(t) &=& y_B (t) \tan\alpha \cos (\omega_W t)\, , \\
y_B(t) &=& y_{W0} + \left[w_B-y_{W0}\right] \frac{t}{t_f}\, .  
\end{array}
\label{wtrajwriggle}
\end{eqnarray}
The box center starts at the right wall of the trap, i.e. $x_B (0) = y_{W0}
\tan\alpha, y_B(0) = y_{W0}$, is then moving down wriggling and ends at $y(t_F)=w_B$, i.e. the box half width.
The resulting trapping fractions $F$ can also be seen in
Figs.~\ref{fig_wedge_3}-\ref{fig_wedge_5} (circles connected by black solid line). The parameters for the
box trajectory have been chosen such that the velocity of the box is limited
$v/\nu < 0.265$ at all times. In the cases $\alpha=30^\circ$ and $\alpha=45^\circ$ we
get an increased fraction of trapped atoms for $t_f/\tau = 40$.

%-----------------------------------------------------------
%-----------------------------------------------------------
%-----------------------------------------------------------

\section{Optimizing box trajectories for a harmonic trap\label{sect4}}

Now, we study the harmonic trap with frequencies $\omega_x = \omega_y = \omega$, see Fig.~\ref{fig_intro_2}b.
The initial state of an atom is chosen concerning the canonical distribution
\begin{eqnarray}
\lefteqn{\rho_{i,H} (x,y,p_x,p_y) = \frac {\omega^2 \beta ^2}{4\pi^2}}&&\nonumber\\
&\times&\exp\left\{- \beta \left[(p_x^2 + p_y^2)/(2m)
+ m \omega^2 (x^2 + y^2)/2 \right]\right\}\, .
\end{eqnarray} 
Again, it is convenient to define a characteristic length $l$, a
characteristic velocity $\nu$ and a characteristic time $\tau$ by
\begin{eqnarray}
l = \frac{1}{\omega}\sqrt{\frac{k_B T_i}{m}},\quad \nu = \sqrt{\frac{k_B T_i}{m}},\quad
\tau = \frac{1}{\omega}.
\end{eqnarray}
In the rest of the paper we are again assuming$~^{87}{\rm Rb}$ atoms, the initial
temperature shall be $T_i = 100\, \mu{\rm K}$ and $\omega= 50 \times 2\pi \, {\rm Hz}$.
The characteristic parameters are then
\begin{eqnarray}
l = 311\, \mu{\rm m},\quad 
\nu = 9.78\, {\rm cm/s},\quad
\tau = 3.18\, {\rm ms}.
\end{eqnarray}

\subsection{Box at rest}

\begin{figure}{}
(a) \includegraphics[width=0.9\linewidth]{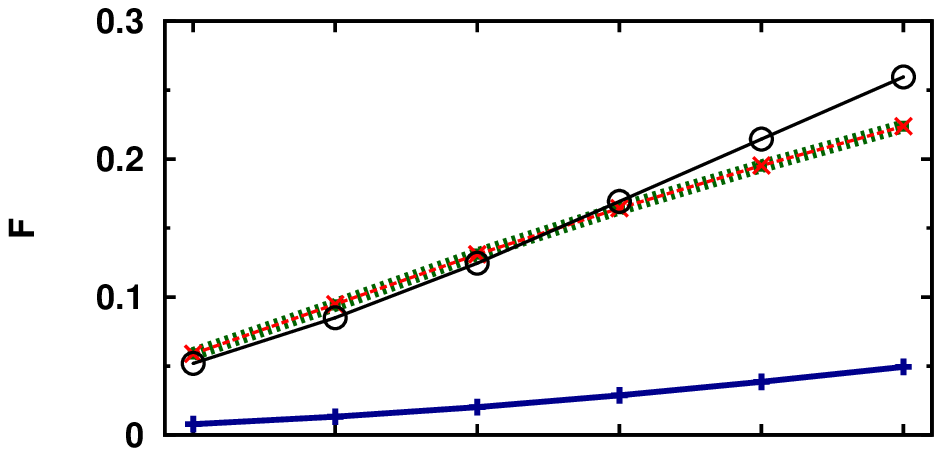}

(b) \includegraphics[width=0.9\linewidth]{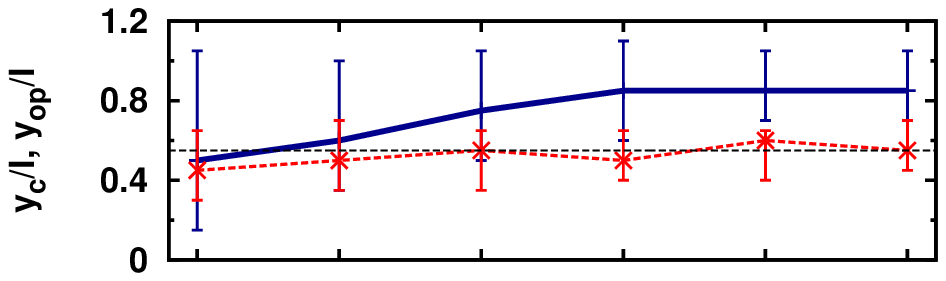}

(c) \includegraphics[width=0.91\linewidth]{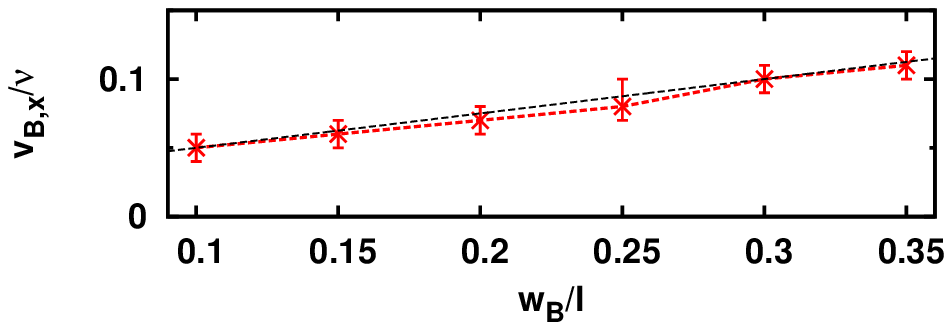}
\caption{(Color online) Harmonic trap, data versus different box half widths $w_B$.
(a) Fraction $F$ of trapped atoms,
box at rest with optimized $y_{op}$ (plus signs connected by thick blue line),
linear moving box with optimized $v_{B,x}$ and optimized $y_{c}$ (crosses
connected by red solid line),
linear moving box with trajectory Eq. (\ref{htrajanal}) (dots connected by
thick green dotted line, on top of red line),
box moving with a helix trajectory (circles connected by black solid line).
(b) Optimized box parameter $y_{op}$ resp. $y_{c}$ (crossing point of $y$-axis):
for a box at rest (plus signs connected by thick blue line),
for a linearly moving box (crosses
connected by red dashed line), approximation Eq. (\ref{happrox1})  (black dashed line).
(c) Optimized box velocity $v_{B,x}$: for a linearly moving box (crosses
connected by red dashed line), approximation Eq. (\ref{happrox2}) (black dashed line).
$E_B/E_i = 0.1, t_f/\tau = 60$.
}
\label{fig_harm_1}
\end{figure}

As a reference, we first look again at the case that the box center is placed at
$x_B=0$ at rest, i.e. $v_{B,x}=v_{B,y}=0$.
The box threshold energy is $E_B/E_i = 0.1$ and the final time is $t_f/\tau = 60$. 
The box coordinate $y_B=y_{op}$ is numerically optimized such that the
resulting fraction of trapped atoms $F$ is maximal.
Fig.~\ref{fig_harm_1}a shows the trapping fraction $F$ for
different box half widths $w_B$ (plus signs connected by thick blue line).
The optimized box coordinate $y_{op}$ is shown in Fig.~\ref{fig_harm_1}b
(plus signs connected by thick blue line). The error bars are defined
by the range in which the maximal trapping fraction $F$ decays by an amount of
$1/\sqrt{N_i}$, where $N_i$ is the number of particles used in the  
numerical simulation. From the error bars, we can see that for small $w_B$ the result is
relatively independent of the exact box position $y_{B}$.

% ---------------- Fig. 9 --------------------------------------
% ---------- Harmonic: Box at rest and linear motion -----------------------------
% ---------------- Fig. 9 --------------------------------------

\subsection{Box with linear motion}

\begin{figure}{}
\begin{center}
\includegraphics[width=0.8\linewidth]{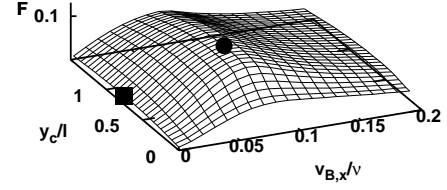}
\end{center}
\caption{Harmonic trap, box moving with the linear trajectory Eq. (\ref{htrajlin}),
fraction $F$ for different positions $y_c$ and box velocities $v_{B,x}$;
$v_{B,y}=0$; $w_B/l = 0.2$; the square symbol marks the maximal fraction for a
box at rest, the dot marks the maximal fraction for a linear moving box.
$E_B/E_i = 0.1, t_f/\tau = 60$.
}
\label{fig_harm_2}
\end{figure}

We want to examine if a linearly moving box can produce a higher fraction $F$ of
trapped atoms in the case of a harmonic trap.
Therefore, we are now looking at a box moving in such a way
that its trajectory crosses the point $x=0,y=y_{c}$ at time $t_f/2$.  
Because of symmetry, we can restrict to the case $v_{B,x}>0, v_{B,y}=0$
(if we neglect the rotation of the box itself).
We consider a linear trajectory of the moving box as
\begin{eqnarray}
\begin{array}{rcl}
x_B(t) &=& v_{B,x} (t-t_f/2)\, , \\
y_B(t) &=& y_c\, .
\end{array}
\label{htrajlin} 
\end{eqnarray}
In Fig.~\ref{fig_harm_2}, the resulting fraction $F$ is shown for the box half
width $w_B/l = 0.2$. The box symbol marks the maximal fraction for a
box at rest while the dot marks the maximal fraction for a linearly moving box.
It can be seen that the linearly moving box traps more atoms than the box at
rest.

We also calculated the fraction for different box half widths, the result can
be seen in Fig.~\ref{fig_harm_1}a.
For every box half width, the
position $y_c$ and the velocity $v_{B,x}$ have been optimized and the optimal
values are shown in Fig.~\ref{fig_harm_1}b and Fig.~\ref{fig_harm_1}c
(crosses connected by red dashed line). The error bars are defined in the same way as explained in
the last subsection.
In Fig.~\ref{fig_harm_1}a it can be seen that in all cases the moving box
traps significantly more atoms than the box at rest.
Good approximations for the optimal position $y_c$ and the velocity $v_{B,x}$ 
are
\begin{eqnarray}
y_c/l &=& 0.55,\label{happrox1}\\
v_{B,x}/\nu &=& 0.025 + 0.25 w_B/l,\label{happrox2}
\end{eqnarray}
which are also shown in 
Fig.~\ref{fig_harm_1}b and Fig.~\ref{fig_harm_1}c (black dashed line). These
approximations lead to an analytical box trajectory of
\begin{eqnarray}
\begin{array}{rcl}
x_B(t) &=& (0.025\,\nu + 0.25 w_B\,/\tau) (t-t_f/2)\, , \\
y_B(t) &=& 0.55\, l\, .
\end{array}
\label{htrajanal} 
\end{eqnarray}
The resulting fraction $F$ for a box moving with trajectory
Eq. (\ref{htrajanal}) is also shown in Fig.~\ref{fig_harm_1}a (dots connected by thick green dotted line on
top of the red line), it is indistinguishable from the result obtained with
trajectory Eq. (\ref{htrajlin}) with optimized parameters $y_c$ and $v_{B,x}$.

% ---------------- Fig. 10 --------------------------------------
% ---------- Harmonic: Linear motion (3D) ----------------------
% ---------------- Fig. 10 --------------------------------------

\begin{figure}{}
\begin{center}
(a) \includegraphics[width=0.8\linewidth]{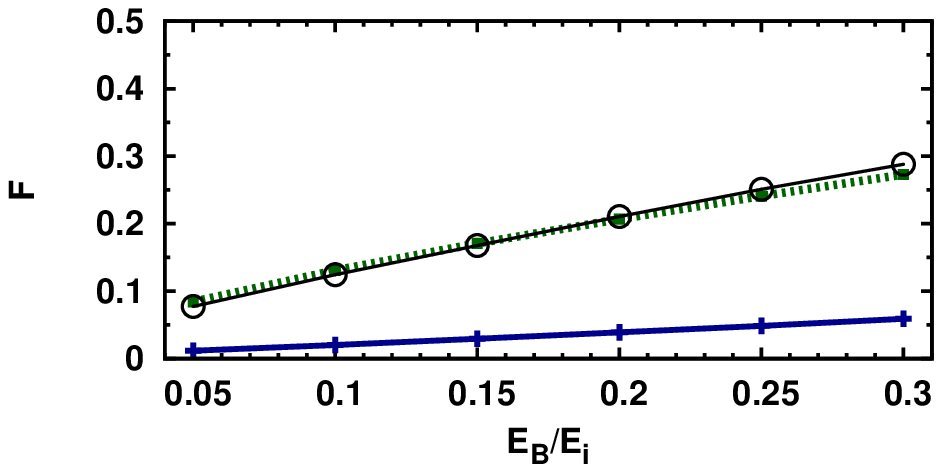}

(b) \includegraphics[width=0.8\linewidth]{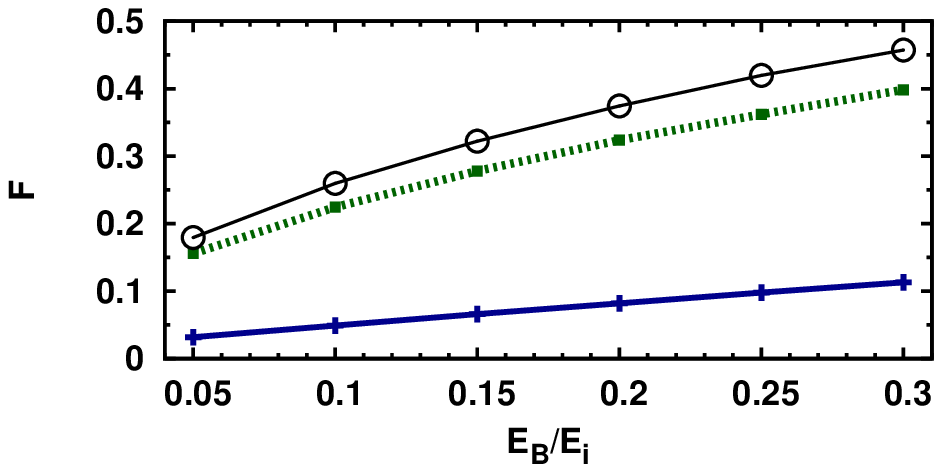}
\end{center}
\caption{(Color online) Harmonic trap, fraction $F$ versus box threshold energies $E_B$:
box at rest (plus signs connected by thick blue line),
box moving with the linear trajectory Eq. (\ref{htrajanal}) (dots connected by
thick green dotted line),
box moving with a helix trajectory (circles
connected by black solid line).
(a) Box half width $w_B/l = 0.2$, (b) box half width $w_B/l = 0.35$.
$t_f/\tau = 60$.}
\label{fig_harm_3}
\end{figure}

In the following we will show that this trajectory Eq. (\ref{htrajanal}) is
also a very good choice for different box threshold energies $E_B$.
The fraction $F$ versus different box threshold energies $E_B$ for the box at rest (plus signs
connected by thick blue line) and the box moving with trajectory Eq. (\ref{htrajanal}) (dots connected by
thick green dotted line) are shown for the box half width $w_B/l = 0.2$ in Fig.~\ref{fig_harm_3}a and
for $w_B/l = 0.35$ in Fig.~\ref{fig_harm_3}b.
It can be seen that even for other  box threshold energies $E_B$ the linearly moving box
traps more atoms in the same time than the box at rest.

% ---------------- Fig. 11 --------------------------------------
% ---------- Harmonic: Linear motion (3D) ----------------------
% ---------------- Fig. 11 --------------------------------------

\subsection{Helix trajectories}

Again, we want to emphasize that the linear trajectory above might not be
optimal but it is probably easier to implement experimentally than a more
complicated motion.

Just as an example, we try a different, more complicated box trajectory. The box
is moving with a helix trajectory given by 
\begin{eqnarray}
\begin{array}{rcl}
x_B(t) &=& x_{H} \left(1- t/t_f \right) \cos (\omega_H t)\, , \\
y_B(t) &=& x_{H} \left(1-t/t_f \right) \sin (\omega_H t)\, ,
\end{array}
\end{eqnarray} 
where $x_{H}$ and $\omega_H$ are the key parameters of the box trajectory.
The results can be seen in Fig.~\ref{fig_harm_1} as well as in
Fig.~\ref{fig_harm_3} (circles connected by black solid line).
The parameters of the helix trajectory are 
$x_{H}/l = 1.9$ and $\omega_H=0.1$.
For larger box half width $w_B$ (see Fig.~\ref{fig_harm_3}b) this trajectory helps to get a larger fraction
of trapped atoms while there is no improvement for smaller box half width.

%-----------------------------------------------------------
%-----------------------------------------------------------
%-----------------------------------------------------------

\section{Summary and Outlook\label{sect5}}
We have examined a model for cooling in a two-dimensional wedge
trap and a two-dimensional harmonic trap. During the cooling procedure,
the atoms are captured by a small area surrounded by ``diodic'' walls, called ``box'', which moves
through the wedge trap and the harmonic trap, respectively. We have examined
different box trajectories with the goal to maximize the fraction of trapped
atoms, this leads also to an increased cooling efficiency. We have shown that
the fraction of caught atoms can be increased using a
moving box compared to a box at rest which was examined in an earlier work
\cite{choi_2010}. We have also optimized the parameters of the box trajectory
where we restricted ourselves mainly to linear box motions due to its possible
easier experimental implementation.
In a future work, we will consider the question of the optimal general box trajectory
in more detail and also taking quantum effects into account.

\section*{Acknowledgments}
VPS thanks H. Wanare, S. Anantha Ramakrishna and R. Vijaya for
fruitful discussions. VPS acknowledges support from the German Academic
Exchange Service under DAAD-IIT Master Sandwich Program.
%
%
% 

%----------------------------------------REFERENCES----------------------------------------------%%----

\end{document}